# A Novel Trajectory Clustering technique for selecting cluster heads in Wireless Sensor Networks

Hazarath Munaga[1], J.V.R. Murthy[1], and N.B.Venkateswarlu[2]
[1] University College of Engineering, Dept. of CSE, J.N.T.University Kakinada, A.P, India
Email: {hazarath.munaga, mjonnalagedda}@gmail.com
[2] AITAM, Dept of CSE, Tekkali, A.P, India
Email: venkat_ritch@yahoo.com

*Abstract*—Wireless sensor networks (WSNs) suffers from the hot spot problem where the sensor nodes closest to the base station are need to relay more packet than the nodes farther away from the base station. Thus, lifetime of sensory network depends on these closest nodes. Clustering methods are used to extend the lifetime of a wireless sensor network. However, current clustering algorithms usually utilize two techniques; selecting cluster heads with more residual energy, and rotating cluster heads periodically to distribute the energy consumption among nodes in each cluster and lengthen the network lifetime. Most of the algorithms use random selection for selecting the cluster heads. Here, we propose a novel trajectory clustering technique for selecting the cluster heads in WSNs. Our algorithm selects the cluster heads based on traffic and rotates periodically. It provides the first trajectory based clustering technique for selecting the cluster heads and to extenuate the hot spot problem by prolonging the network lifetime.

*Index Terms*—Trajectory clustering, Wireless sensor networks, Network life time, Cluster head

## I. INTRODUCTION

Wireless sensor networks (hereinafter, WSNs) are networks of wireless nodes that are deployed over an area for the purpose of monitoring certain phenomena of interest. To keep specific areas under observation, WSNs deploy hundreds or thousands of integrated sensor nodes to sample data from observed environment. The nodes perform certain measurements, process the measured data and transmit the processed data to a base station over a wireless channel. The base station collects data from all the nodes, and analyzes this data to draw conclusions about the activity in the area of interest. In practice, due to the large quantity of sensor nodes, it is infeasible to recharge the batteries in WSNs. Therefore, sensor network lifetime is a primary concern in sensor network design.

In literature many researchers concerning protocols for WSNs have been proposed to improve the energy consumption and the network lifetime. Those protocols can be categorized into three classes: routing protocols, sleep-and-awake scheduling protocols, and clustering protocols. The routing protocols [1] [2] determine the energy-efficient multi-hop paths from each node to the base station. In sleep-and-awake scheduling protocols [3-5], every node in the schedule can sleep, in order to minimize energy consumption. In clustering protocols [6] [7] data aggregation can be used for reducing energy consumption. Data aggregation, also known as data fusion, can combine multiple data packets received from different sensor nodes. It reduces the size of the data packet by eliminating the redundancy. Wireless communication cost is also decreased by the reduction in the data packets [8]. Therefore, clustering protocols improve the energy consumption and the network lifetime of the WSNs.

Clustering [9] is a commonly adopted approach in sensor networks to manage power efficiently. In clustering, sensors in the monitoring area are grouped into clusters; all sensor nodes within the same cluster send their data to the cluster head, which then forwards the aggregated data to the base station. Therefore, cluster heads "typically die at an early stage" [10]. This is sometimes called as the hot spot problem [11]. Without adding extra nodes or redistributing the available energy, this problem is hard to solve. For example, [10] have shown that varying the transmission power of nodes, even considering unlimited transmission ranges, does not solve the hot spot problem. At the same time, it is also envisioned that sensor nodes will become "extremely inexpensive" [12]. While beyond a certain node density, adding additional nodes does not provide any improvement regarding sensing, communication or coverage [13], adding nodes might obviously help to increase the lifetime of a sensor network while providing the same service to its users, i.e. leveraging sensor values from the same number of nodes.

Ref. [14] proposed LEACH, a well-known clustering protocol for WSNs. LEACH includes distributed cluster formation, local processing to reduce global communication and randomized rotation of cluster heads among all the nodes in the network. Each cluster selects a cluster head, which is responsible for aggregating collected data and sending data to base station. LEACH provides a good model that helped to reduce information overload and provides a reliable data to the end user. Together, these features allow LEACH to achieve the desired properties.

Ref. [15] the problem of finding an energy-balanced solution to data propagation in WSNs using a probabilistic algorithm was considered for the first time. The lifespan of the network is maximized by ensuring that the energy consumption in each slice is the same. Sensors are assumed to be randomly distributed with uniform distribution in a circular region or, more generally, the sector of a disk. Data have to be propagated by the WSN towards a sink located at the center of the disk, and it is shown that energy balance can be achieved if a recurrence relation between the probabilities that a slice ejects a message to the sink is satisfied.

Ref. [16] proposed clustering-based routing protocol called base station controlled dynamic clustering protocol (BCDCP), which utilizes a high energy base station to set up cluster heads and perform other energy-intensive tasks, can noticeably enhance the lifetime of a network.

Ref. [17] proposed two new algorithms under the name PEDAP, which are near optimal minimum spanning tree based wireless routing scheme. The performance of the PEDAP was compared with LEACH and PEGASIS, and showed a slightly better network lifetime than PEGASIS. Ref. [18] proposed a new routing scheme (SHORT), to achieve higher energy efficiency, network lifetime, and more throughput than PEGASIS, and PEDAP-PA protocols. This scheme used the centralized algorithms and required the powerful base station. The performance results showed that SHORT can achieve better "energy X delay" performance than the existing chain based data aggregation protocols.

Ref. [19] proposed EECR, which is an energy efficient clustering routing algorithm. The performance of the EECR was compared with LEACH, and showed a slightly better network lifetime than LEACH.

However, the unsolved problem of considerable energy consumption on the cluster formation still exists. Here, we consider the path followed by the node to transfer data to the base station as the *"trajectory"*. We used our proposed novel trajectory clustering algorithm for clustering such paths and obtained *"representative trajectory"* is used to assign the cluster heads. These obtained cluster heads will be used for communicating data to the base station. In this paper, we concentrated on the rotation of cluster heads among all sensor nodes to improve the lifetime of the network based on the traffic density. We tested our proposed method and found that this method enhances the lifetime of the network.

## II. NOVEL ALGORITHM

This section considers the WSNs consisting of hundreds or thousands of deployed sensor nodes in the sensing field. On the basis of [20][16], it is assumed by the following properties of the WSNs to simplify the network model.

- The base station is located far away from the sensors,
- The nodes have uniform initial energy allocation and all sensor nodes have equal capabilities (data processing, wireless communication, battery power).
- All sensor nodes have various transmission power levels, and each node can change the power level dynamically.
- Each node senses the environment at a fixed rate, and
- All nodes are immobile.

The sensor nodes are geographically grouped into clusters and capable of operating in two basic modes: the sensing mode and the cluster head mode [20]. In the sensing mode, the node senses the task and sends the sensed data to its cluster head. In cluster head mode, a node gathers data from its cluster members, performs data fusion, and transmits the data to the base station. The base station in turn performs the key task of cluster head selection.

### A. Cluster Head Selection

Initially the nodes will transmit a hello packet to the base station. After receiving hello packets from the nodes, using the Trajectory Clustering algorithm, the base station computes the representative trajectory by clustering the trajectories (here the trajectory is nothing but the path used by the node to transfer its data to the base station). The nodes of the obtained representative trajectory are considered as the cluster heads. Then the base station splits the network into clusters (equal to the number of nodes in the representative trajectory), and identifies the nodes in the representative trajectory as the corresponding cluster heads. Then, the base station broadcast a message to the network mentioning about the nodes and their corresponding cluster heads. Subsequently the nodes will use its cluster heads to transmit data. This process will be performed periodically and the cluster heads will change based on the traffic.

Cluster head selection routine contains the following stages:-
1. Base station computes the cluster heads using proposed Trajectory Clustering algorithm;
2. Split the network into N clusters; and
3. Broadcast message to all nodes mentioning cluster members and their corresponding cluster heads

### B. Trajectory Clustering

The success of any clustering algorithm depends on the adopted dissimilarity measure. Following section explains about the adopted dissimilarity measure.

Ref. [21], proposed the usage of Euclidean distance between time series of equal length as the measure of their similarity. The idea has been generalized in [22] for subsequence matching. In a similar way [23] used Discrete Wavelet Transform and [24] used Principal Component Analysis for measuring time series similarity. Another approach which is brought from image processing is *Time Warping technique* and it is used in [25] to match signals in speech recognition. A similar technique is used to find longest common subsequence (LCSS) of two sequences using fast probabilistic algorithms to compute the LCSS, and then define the distance using the length of this subsequence [26].

Here we adopted Hausdorff distance [27] for calculating dissimilarity between trajectories. The following are some of the definitions used in our algorithm.

*Definition 1*: A trajectory (t) is represented as $trj(t_{id}, u_0, u_1, u_2, ..., u_n)$ where $(t_{id})$ is a unique trajectory id (data packet), and $(u_0, u_1, u_2, ..., u_n)$ is a sequence of nodes reflecting the spatial position of the node.

*Definition 2*: We define the spatial dissimilarity function between two trajectories $t_1$ and $t_2$ as the maximum of one way distances between two trajectories. The one way distance from a trajectory $t_1$ to another trajectory $t_2$ is defined as the integral of the Hausdorff distance between points of $t_1$ to trajectory $t_2$ divided by the number of points in $t_1$ ($/t_1/$).

$$dist_{ow}(t_1, t_2) = \frac{1}{|t_1|} \int_{p \in t_1} d_h(p, t_2) dp$$

The Hausdorff distance from a trajectory point $p$ to another trajectory $t_2$ is defined as $d(p, t_2) = \min_{q \in t_2} \{d(p,q)\}$. The distance between trajectories $t_1$ and $t_2$ is the maximum of their one way distances, i.e., $dist(t_1, t_2) = \max\{dist_{ow}(t_1, t_2), dist_{ow}(t_2, t_1)\}$

Clearly the $dist_{ow}(t_1, t_2)$ is not symmetric but $dist(t_1, t_2)$ is symmetric. Note that $dist_{ow}(t_1, t_2)$ is the integral of the shortest distances from points in $t_1$ and $t_2$.

*1) Trajectory Cluster Routine*

Trajectories are grouped into clusters using the *threshold*. Here the threshold is considered as a maximum value, such that all trajectories are grouped into a single cluster. The trajectory cluster routine contains the following stages:

1. Dissimilarity matrix for trajectories will be computed using the Hausdorff distance,
2. Using following *Initialization* Algorithm trajectories are grouped into initial clusters;
   a. Take first sample as first cluster. Classify all the remaining trajectories into this cluster if they are within the threshold.
   b. Take a trajectory (sequentially) which is not already classified into any of the cluster and consider it as a new cluster. Take all the other trajectories which are not kept in any of the clusters and keep in this cluster if they satisfy the threshold limit.
   c. Repeat step b till no new cluster is added.
3. Using the following *RepTraj* Algorithm representative trajectories are computed.
   a. For each Trajectory of cluster C calculate cumulative dissimilarity with all other trajectories of the same cluster C. Select the trajectory which is having minimum cumulative dissimilarity and take this as representative trajectory of that cluster.
4. By considering the trajectories received from step 3, as initial cluster canters, using the following *Re-cluster* Algorithm re compute clusters and their representative trajectories until there is no change in the representative trajectories.
   a. For each Trajectory calculate dissimilarity with all the K representative trajectories and classify to the cluster for which dissimilarity is low.
   b. Re-calculate representative trajectories using *RepTraj* Algorithm.

*C. Data communication phase*

There are three steps during the data communication phase: data collection, data fusion and data transmission. Initially each sensor node transmits the sensed information to its cluster head at the time slot assigned by its cluster head. In order to save itself energy, the node will close transmit part during the time slot, which is not required to it. Once data from all sensor nodes have been received, the cluster head performs data fusion on the collected data and reduces the amount of raw data that need to send to the base station. Once the data gathering and data fusion are completed, the cluster head sends the compressed data to the base station.

As mention previously, all the nodes can work as a cluster heads. Due to this, any node can become a cluster head or a cluster member. At each turn the cluster head calculates available power and compares with the cluster members. Whenever the cluster heads power becomes less than the minimum power holding, then the cluster heads informs to its cluster members and assigns the maximum power holding cluster member as the cluster head and in turn communicates to the base station. Whenever the cluster head is changed base station repeats the cluster finding process and modifies the clusters.

III. EXPERIMENTAL WORK

To evaluate the performance of our algorithm, it has been simulated and compared its performance with energy efficient clustering routing (hereafter, EECR). Before the simulation and results are introduced, the radio model and some important parameters [19] used in simulation have been described.

*A. The radio model*

We have used both the free-space propagation model and the two-ray ground propagation model to approximate path loss sustained because of wireless channel transmission. Given a threshold transmission distance of $d_0$, the free-space model is used when $d < d_0$, and the two-ray model is applied for cases when $d \geq d_0$. Using these two models, the transmit energy costs for the transfer of a b-bit data message between two nodes separated by a distance of d meters is given:

$$\text{if } d < d_0, \ E_T(b,d) = E_{Tx}b + E_{amp}(d)b = E_{Tx}b + \varepsilon_1 d^2 b \quad (1)$$

$$\text{if } d \geq d_0, \ E_T(b,d) = bE_{BF} + \varepsilon_2 d^4 b \quad (2)$$

With regard to the energy cost incurred in the receiver of the destination node, we give in Eq. (3):

$$E_T(b) = E_{Rx}b \quad (3)$$

We have summarized the different meanings and values for energy terms in Table 1. Energy consumed during data aggregation in the cluster head $E_{da}$, is also taken into account.

TABLE I.
SUMMARIZES MEANING OF EACH TERM AND TYPICAL VALUE

| Term | Meaning | Value |
|---|---|---|
| $E_{da}$ | Consume energy for data aggregation | 5nJ/bit |
| $E_{Tx}$, $E_{Rx}$ | Radio Electronics Energy | 50nJ/bit |
| $\varepsilon_1$ | Transmit applied for free space | 10pJ/(bit*m$^2$) |
| $\varepsilon_2$ | Transmit applied for two way model | 0.0013 pJ/(bit*m$^4$) |

### B. The number of clusters

We assume that N nodes are distributed in the area of A*A randomly. If there are *M* clusters, then there are *N/M* nodes in each cluster on an average. Every cluster head receives the sensed data from its cluster nodes, aggregates all the data, and sends it to the base station. The total energy spent on transmitting a frame for every cluster head can be expressed as:

$$E_1 = bE_{Tx}\frac{N}{M} + bE_{da}\frac{N}{M} + b\varepsilon_2 d_1^4 \qquad (4)$$

where $d_1$ is the distance between cluster head and the base station.

In one frame, the cluster nodes transmit the sensed data messages to its cluster head. The energy spent for each cluster member is as below:

$$E_2 = bE_{Tx} + b\varepsilon_1 d_2^2 \qquad (5)$$

where $d_2$ is the distance between the member node and its cluster head. If the cluster head is in the centre of the cluster, the density of every cluster is $\rho = M/A^2$, then $d_2$ can equate to

$$d_2 = \sqrt{\frac{1}{2\pi}\frac{A^2}{M}} \qquad (6)$$

The energy spent for each cluster member is modified as:

$$E_2 = bE_{Tx} + b\varepsilon_1 \frac{1}{2\pi}\frac{A^2}{M} \qquad (7)$$

The energy dissipation in a cluster can be expressed as:

$$E_c = E_1 + \left(\frac{N}{M}-1\right)E_2 \qquad (8)$$

The total energies dissipated in all the clusters can be expressed as:

$$E = ME_c = b\left[\begin{array}{c} 2E_{Tx}N + E_{da}N + M\varepsilon_2 d_1^4 \\ + (N-M)\varepsilon_1 \frac{1}{2\pi}\frac{A^2}{M} \end{array}\right] \qquad (9)$$

If $\frac{\partial E}{\partial M} = 0$, we can get the following Eq. (10)

$$M = A\sqrt{\frac{N}{2\pi}\frac{\varepsilon_1}{(\varepsilon_2 d_1^4 - E_{Tx})}} \qquad (10)$$

In our simulation, we consider $N = 100$, $A = 100$ m and $d_1 = 90$, and for various number of clusters i.e., from six to twelve.

*1) Results*

It has been simulated that 100 nodes randomly located in the sensing field of 100 X 100 m$^2$ with the base station located at least 90 m away. All sensor nodes periodically sense events and transmit the data packet to the base station. All sensor nodes start with an initial energy of 2 J and the data message size is fixed at 516 bytes, of which 16 bytes represent the weight value. We choose three different coefficients $C_1 = 0.5$, $C_2 = 0.4$ and $C_3 = 0.1$. To evaluate the performance of our algorithm, we compare its performance with EECR.

Performance is measured by the number of rounds alive and the total data messages successfully delivered. As shown in Fig. 1 and 2, it's proved that our algorithm outperforms EECR and LEACH in the following way:

1. *In the number of rounds the nodes alive*: Ref. [19] shown that EECR exceeds LEACH by more than 45 % when the number of rounds is above 100. The nodes that remain alive in EECR are a maximum of 175 rounds, whereas with our proposed method rounds alive are 350 (*see* Fig. 1).
2. *Number of packets delivered*: If the system life time is defined as the number of rounds alive, with our proposed technique system life can increase 90%. Subsequently the number of packets delivered at the base station during the number of rounds of activity is increased from a maximum of 40000 to 70000 (*see* Fig. 2).

Hence, from the above analysis, it is found that our algorithm can achieve lower dissipation value of energy, higher data messages delivery, and effectively postpone the system lifetime than those of EECR and LEACH.

### IV. CONCLUSION

In this paper, a novel trajectory based clustering solution is presented for selecting cluster heads in WSNs. Trajectory clustering algorithm enables sensor nodes to reduce data packets by data aggregation. The wireless communication cost is decreased by reduction of data packets, and thus the clustering technique extends the lifetime by reducing the energy consumption of the network. The simulation results demonstrated that our proposal significantly improves the lifetime and reduce the energy consumption of WSNs compared with existing clustering protocols. We assume that the nodes are error free. However, error will arise due to the noise in the real network environments. As a future work, we plan to extend the method to increase its robustness.

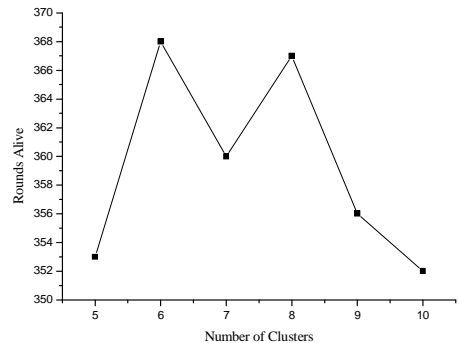

Figure 1. Number of rounds alive

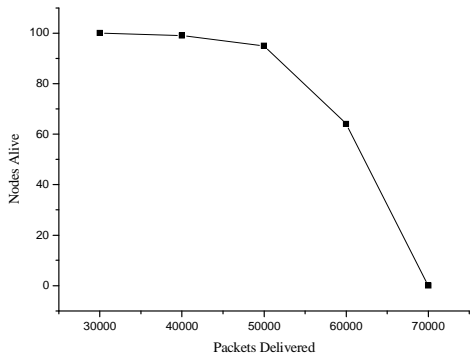

Figure 2. Packets delivered

REFERENCES

[1] R. Shah and J. Rabaey, "Energy aware routing for low energy adhoc sensor networks," in *WCNC2002: Wireless Communications and Networking Conference,* vol. 1. Washington, DC, USA: IEEE Computer Society, pp. 350–355, March 2002.

[2] J.-H. Chang and L. Tassiulas, "Maximum lifetime routing in wireless sensor networks," *IEEE/ACM Trans. Netw.,* vol. 12, no. 4, pp. 609–619, Aug. 2004.

[3] F. Ye, G. Zhong, S. Lu, and L. Zhang, "Peas: A robust energy conserving protocol for long-lived sensor networks," in *10th IEEE International Conference on Network Protocols*, pp. 200–201, 2002.

[4] C. Gui and P. Mohapatra, "Power conservation and quality of surveillance in target tracking sensor networks," in *MobiCom '04: International conference on Mobile computing and networking*, New York, USA: ACM, pp. 129–143, 2004.

[5] J. Deng, Y. S.Han, W. Heinzelman, and P. Varshney, "Balanced-energy sleep scheduling scheme for high density cluster-based sensor networks," in *ASWN 2004: 4th Workshop on Applications and Services in Wireless Networks*, pp. 99–108, 2004.

[6] C. Li, M. Ye, G. Chen, and J. Wu, "An energy-efficient unequal clustering mechanism for wireless sensor networks," in *IEEE International Conference on Mobile Adhoc and Sensor Systems Conference*, p. 604, 2005.

[7] C.-Y. Wen and W. A. Sethares, "Automatic decentralized clustering for wireless sensor networks," *EURASIP J. Wirel. Commun. Netw.*, vol. 5, no. 5, pp. 686–697, Oct. 2005.

[8] B. Krishnamachari, D. Estrin, and S. B. Wicker, "The impact of data aggregation in wireless sensor networks," in *ICDCSW '02: 22nd International Conference on Distributed Computing Systems*. Washington, DC, USA: IEEE Computer Society, pp. 575–578, 2002.

[9] T. J. Kwon and M. Gerla, "Clustering with power control," in *MILCOM 1999: IEEE Military Communications Conference*, vol. 2. IEEE Computer Society, pp. 1424–1428, 1999.

[10] M. Perillo, Z. Cheng, and W. Heinzelman, "On the problem of unbalanced load distribution in wireless sensor networks," in *IEEE Global Telecommunications Conference Workshops*, pp. 74–79, 2004.

[11] S. Ahn and D. Kim, "Proactive context-aware sensor networks," in *EWSN 2006: Third European Workshop*, vol. 3868. Springer Berlin/ Heidelberg, pp. 38–53, 2006.

[12] D. Estrin, R. Govindan, J. Heidemann, and S. Kumar, "Next century challenges: scalable coordination in sensor networks," in *MobiCom '99: 5th annual ACM/IEEE international conference on Mobile computing and networking*. New York, USA: ACM, pp. 263–270, 1999.

[13] L. G. Nirupama Bulusu, Deborah Estrin and J. Heidemann, "Scalable coordination for wireless sensor networks: Self-configuring localization systems," in *ISCTA 2001: Sixth Int. Symp. on Comm. Theory and App.*, pp. 1–6. July 2001.

[14] W. R. Heinzelman, A. Chandrakasan, and H. Balakrishnan, "Energy efficient communication protocol for wireless micro sensor networks," in *HICSS '00: 33rd International Conference on System Sciences,* vol. 8. Washington, DC, USA: IEEE Computer Society, p. 8020, 2000.

[15] C. Efthymiou, S. Nikoletseas, and J. Rolim, "Energy balanced data propagation in wireless sensor networks," Parallel and Dist. Processing Symp. vol. 13, p. 225a, 2004.

[16] S. Muruganathan, D. Ma, R. Bhasin, and A. Fapojuwo, "A centralized energy-efficient routing protocol for wireless sensor networks," *IEEE Communications Magazine*, vol. 43, no. 3, pp. 8–13, March 2005.

[17] H.O. Tan and I. Korpeoglu, "Power efficient data gathering and aggregation in wireless sensor networks," *SIGMOD Rec.*, vol. 32, no. 4, pp. 66–71, Dec. 2003.

[18] Y. Yang, H. Wu, and H. Chen, "Short: shortest hop routing tree for wireless sensor networks," *Int. J. Sen. Netw.*, vol. 2, no. 5/6, pp. 368–374, July 2007.

[19] L. Li, D. Shu-song, and W. Xiang-ming, "An energy efficient clustering routing algorithm for wireless sensor networks," *China Universities of Posts and Telecommunications*, vol. 13, no. 3, pp. 71–75, Sept. 2006.

[20] S. Lindsey, C. Raghavendra, and K. M. Sivalingam, "Data gathering algorithms in sensor networks using energy metrics," *IEEE Trans. Parallel Distrib. Syst.*, vol. 13, no. 9, pp. 924–935, Sept. 2002.

[21] R. Agrawal, C. Faloutsos, and A. N. Swami, "Efficient Similarity Search In Sequence Databases," in *FODO: 4th International Conference of Foundations of Data Org. and Algo.*, Springer Verlag, pp. 69–84, 1993.

[22] C. Faloutsos, M. Ranganathan, and Y. Manolopoulos, "Fast subsequence matching in time-series databases," in *SIGMOD '94: International conference on Management of data*, New York, USA: ACM, pp. 419–429, 1994.

[23] Z. R. Struzik and A. Siebes, "Measuring time series' similarity through large singular features revealed with wavelet transformation," in *DEXA '99: 10th International Workshop on Database and Expert Systems Applications*, Washington, USA: IEEE Computer Society, p. 162, 1999.

[24] M. Gavrilov, D. Anguelov, P. Indyk, and R. Motwani, "Mining the stock market (extended abstract): which measure is best?" in *KDD '00: Sixth Int. conference on KDDM*, New York, USA: ACM, pp. 487–496, 2000.

[25] H. Sakoe and S. Chiba, *Readings in speech recognition* Morgan Kaufmann Publishers Inc., pp. 159 – 165, 1990.

[26] B. Bollobas, G. Das, D. Gunopulos, and H. Mannila, "Time-series similarity problems and well-separated geometric sets," in *International Symposium on Computational Geometry*, pp. 454–456, 1997.

[27] D. P. Huttenlocher and K. Kedem, "Computing the minimum hausdorff distance for point sets under translation," in *SCG '90: sixth annual symposium on Comp. geo*. New York, USA: ACM, pp. 340–349, 1990.